\documentclass[11pt,letterpaper]{JHEP3}
\usepackage{graphicx}
\usepackage{amsfonts}
\usepackage{amssymb}
\usepackage{amsmath}
\usepackage{eucal}
\usepackage{epsfig}

\newcommand{\be}{\begin{equation}}
\newcommand{\ee}{\end{equation}}
\newcommand{\bea}{\begin{eqnarray}}
\newcommand{\eea}{\end{eqnarray}}

\title{Not $T$-parity but $\EuScript{C}$-parity}
\author{Jae~Yong~Lee\\
School of physics, Korea Institute for Advanced Study,\\
Hoegiro 87, Dongdaemun-Gu, Seoul 130-722, Korea\\
E-mail: \email{littlehigg@kias.re.kr}}
\preprint{KIAS-P07100}
\abstract{We revisit the Littlest Higgs model with $T$-parity,
and discover a $Z_2$ symmetry on collective symmetry. 
It is dubbed collective parity ($\EuScript{C}$-parity).
We demonstrate that $T$-parity is consistent with $\EuScript{C}$-parity.
We further investigate the origin of
the collective symmetry in the context of composite Higgs and
find a new path to the UV complete theory of the model.
In addition, we demonstrate that $T$-parity
violating processes naturally take place.}  

\begin{document}
\section{Introduction}
The Large Hadron Collider will begin to operate in 2008 and
is expected to disclose the mechanism for electroweak symmetry breaking (EWSB).
In the theory side, there are already various mechanisms for EWSB on the market.
In this paper we revisit the
``little Higgs" mechanism~\cite{ArkaniHamed:2001nc,Arkani-Hamed:2002qy}
which is based upon the nonlinear sigma models. Especially we
concentrate on the Littlest Higgs model (LHM) which has been rigorously analyzed. 
In the LHM, the SM Higgs is a pseudo Nambu-Goldstone boson (pNGB)
which arises from some global symmetry breaking and its mass is protected
against the infamous quadratic divergence by the so-called
``collective symmetry breaking", which is naturally realized
by doubling of certain particles with the same statistics.
On top of that, the notion of collective symmetry breaking is extremely engineered
and hence leads to the  ``twin Higgs" theories~\cite{Chacko:2005pe,Chacko:2005un}
which explicitly exhibit a discrete $Z_2$ symmetry.

Similar to other mechanisms there is a downside to the little Higgs mechanism:
in order to be compatible with the electroweak precision tests (EWPTs)
the global symmetry breaking scale has to become $\cal{O}$(10) TeV.
So the fine-tuning of the Higgs mass arises once again.
To avoid this problem a new $Z_2$ symmetry between particles of the same statistics
was introduced in the similar way as $R$-parity 
in the Minimal Supersymmetric Stadard Model~(MSSM) where $R$-parity distinguishes
between the SM particle and its superpartner.
In little Higgs theories, a discrete $Z_2$ symmetry is dubbed
``$T$-parity"~\cite{Cheng:2004yc,Low:2004xc}.
In the twin Higgs model, a $Z_2$ symmetry is in fact a principal guide
to construct the model itself.
However, $T$-parity in the Little Higgs theories is given by hand
such as heavy particles are $T$-odd and light particles are $T$-even.

As payback for the introduction of $T$-parity to little Higgs theories,
the number of quarks and leptons in the LHT at least double that of the SM and
there are two kinds of quarks and leptons:
the $T$-even particles are the SM particles while the $T$-odd particles
will give rise to testable new physics at the TeV scale.
Moreover, due to invariance under $T$-parity, the neutral lightest $T$-odd parity
particle (LTOP) is stable and could be a candidate for dark
matter~(DM)~\cite{Hubisz:2004ft,Hubisz:2005bd}
just as the lightest supersymmetric particle (LSP) of the MSSM. 
But recently there was an assertion that $T$-parity in the LHM could be
violated by anomalies so that the LTOP, the heavy photon, decays into
$ZZ$ and $W^+W^-$~\cite{Hill:2007nz,Hill:2007zv} and cannot be a good DM
candidate. This argument is based on the assumption that
the pNGBs are composite particles and the underlying UV physics
is technicolor-like strong interactions.

On the other hand, the composite little Higgs model~\cite{Katz:2003sn,Lee:2005kd}
was independently explored as a natural UV complete theory of the LHM.
It promises the so-called ``dark matter parity" for
some composite fermions arising from the strong interaction sector,
and thus provides a {\it fermionic} DM candidate.
In this light, this discrete symmetry is clearly irrelevant to $T$-parity
so that a UV complete theory of the LHT which is based on composite Higgs
may embrace a new DM candidate. 
Therefore it is challenging to explicitly construct the LHT
in the context of composite Higgs. However, we will not
build a completely new model. Rather, we revisit $T$-parity, 
searching for a $Z_2$ symmetry within the composite Higgs model. 
We find a path not only to the UV completion of the LHT
but also to a good candidate for DM.
 
This paper is organized as follows.
In Section 2, we review the LHT. We thoroughly examine the implementation 
of $T$-parity on the Higgs sector, gauge and fermion sectors. 
In Section 3, we calculate the effective Higgs potential in light of $T$-parity.
We explain why the Higgs potential has no renormalizable interactions
between the Higgs doublet and triplet. Then we identify a discrete 
$Z_2$ symmetry with the invariance under the interchange between
the collective symmetry. This $Z_2$ symmetry is dubbed $\EuScript{C}$-parity. 
In Section 4, we search for the origin of $\EuScript{C}$-parity
by exploiting the composite Higgs.
In Section 5, we discuss the anomaly conditions for the model,
and look for the UV complete theory of the model.
In Section 6, we summarize this paper.

\section{Littlest Higgs model with $T$-parity}
The Littlest Higgs model (LHM) was based on an non-linear sigma model in which
a global $SU(5)$ symmetry is spontaneously broken to $SO(5)$ at a scale $f$
via a VEV of an $SU(5)$ tensor $\Sigma$ field,
\be\label{eq:vev5}
\Sigma_0=\left(\begin{array}{ccccc}  
\,\,0\,\, & 0\,\, & 0\,\, & 1\,\, & 0\,\,\\
\,\,0\,\, & 0\,\, & 0\,\, & 0\,\, & 1\,\,\\
\,\,0\,\, & 0\,\, & 1\,\, & 0\,\, & 0\,\,\\
\,\,1\,\, & 0\,\, & 0\,\, & 0\,\, & 0\,\,\\
\,\,0\,\, & 1\,\, & 0\,\, & 0\,\, & 0\,\,\end{array}\right).
\ee
The $\Sigma$ field transforms under the global $SU(5)$ rotation $V$ as
\be\label{eq:Vtrans}
\Sigma(x)\to V\Sigma(x)V^T.
\ee
Note that the upper (lower) $2\times 2$ of $\Sigma$ is invariant under the 
$SU(3)$ in the lower (upper) block of $V$. This is the most important property
so we will refer to it again and again in later sections.

The global symmetry breaking yields fourteen Nambu Goldstone bosons (NGBs),
among which four NGBs are eaten to become the longitudinal modes
of the heavy partners of the SM $SU(2)_L\times U(1)_Y$ gauge fields.
The ten remaining NGBs are parameterized by the non-linear sigma tensor field,
\be\label{eq:sigmatrans}
\Sigma(x)=\xi(x)\Sigma_0\xi^T(x)=\xi^2(x)\Sigma_0,
\ee
where $\xi(x)=\exp\{i\Pi(x)/f\}$, and $\Pi(x)$ is a $5\times5$ matrix of the NGB fields,
\be
\Pi=\left(\begin{array}{ccc} 0\,\, & H^T/\sqrt{2} \,\,& \Phi \\
H^\ast/\sqrt{2} \,\,& 0\,\, & H/\sqrt{2} \\
\Phi^\dagger\,\, & H^\dagger/\sqrt{2}\,\, & 0\end{array}\right).
\ee
The $H$ field contains four degrees of freedom while
the $\Phi$ field includes six degrees of freedom.
So $H$ and $\Phi$ are the doublet and triplet, respectively, under the SM $SU(2)_L$.
The component fields of $H$ are $G^{\pm,0}$ and $h^0$. $G^{\pm,0}$
are eaten by the SM $SU(2)_L$ gauge bosons while $h^0$ becomes the physical SM Higgs.
The component fields of $\Phi$ consist of five physical scalars
($\phi^{\pm\pm},\phi^\pm,\phi^0$) and a physical pseudo-scalar $\phi^p$.
To leading order, all the physical triplet states have degenerate masses
with an order of $f$. In the LHM the components of $H$ and $\Phi$ mix due to EWSB
while they do not in the LHT.
From Eq.~(\ref{eq:Vtrans}) and (\ref{eq:sigmatrans}) 
we can infer that $\xi$ transforms under the $SU(5)$ as
\be
\xi(x)\to U\xi(x)\Sigma_0V^T\Sigma_0=V\xi(x)U^\dagger.
\ee
$U$ takes values in the Lie algebra of the unbroken $SO(5)$ subgroup, 
and is a nonlinear function of both the $V$ matrix and $\Pi(x)$.
The transformation of $\xi$ field is indispensable to describe heavy fermions
later.

\subsection{$T$-parity and Higgs sector}
Followed by the terminology of Cheng and Low~\cite{Cheng:2004yc,Low:2004xc},
$T$-parity is implemented to the NGB matrix $\Pi$ as follows:
\be\label{eq:tahiggs}
T\,:\,\,\Pi\to -\Omega \Pi\Omega
\ee
with $\Omega= \mbox{diag }(1,1,-1,1,1)$, implying that the doublet Higgs $H$
is $T$-even and  the triplet $\Phi$ is $T$-odd:
\be\label{eq:tahiggs2}
T\,:\,\, \left\{\begin{array}{r@{\quad\rightarrow\quad}l} 
H & H \\ \Phi&-\Phi. \end{array}\right.
\ee
$T$-parity differentiates between the doublet and triplet so
no mixing between their components occurs at electroweak scale
in contrast with the Littlest Higgs model without $T$-parity.
Therefore $T$-parity is a promising feature of the LHM
to avoid fine-tuning of Higgs mass parameter. 
However it seems {\it unnatural} to some extent that the components
of a single multiplet under the global $SU(5)$ have {\it different} parities.
It implies that $T$-parity does not respect the $SU(5)$ symmetry.
A certain discrete symmetry must live in the subgroups of the $SU(5)$
so that $T$-parity may be derived as an accidental symmetry from
this discrete symmetry. 

For later use we set out the transformations of $\xi$ and $\Sigma$, respectively,
under $T$-parity:
\be
T\,:\,\,\xi\to \Omega\xi^\dagger\Omega,
\ee
\be
T\,:\,\, \Sigma \to \tilde\Sigma=\Sigma_0\Omega\Sigma^\dagger\Omega\Sigma_0,
\ee
where $\tilde\Sigma$ is the dual of $\Sigma$ which is obtained 
by substituting $\Phi$ in $\Sigma$ for $-\Phi$.
 
\subsection{$T$-parity and gauge boson sector}
The mechanism of collective symmetry breaking drives the LHM to possess
a pair of $SU(2)\times U(1)$ gauge symmetries which are broken down
to the diagonal subgroup identified with the SM $SU(2)_L\times U(1)_Y$.
$T$-parity is also introduced to the gauge sector
in order to decouple the broken gauge fields to the SM fields.
The implementation of $T$-parity on the gauge fields is to interchange the pair 
of gauge bosons:
\be\label{eq:tagauge}
T\,:\,\,W^a_1\leftrightarrow W^a_2,\qquad
B_1\leftrightarrow B_2,
\ee
where $W^a_{1,2}$ and $B_{1,2}$ are gauge fields of $SU(2)_{1,2}$ and $U(1)_{1,2}$, respectively.
The invariance under the interchange is allowed only if the gauge couplings
meet the conditions
\be\label{eq:gaugecoupling}
g_1=g_2=\sqrt{2}g,\qquad g^\prime_1=g'_2=\sqrt{2}g',
\ee
where $g$ and $g'$ are the coupling constant of the $SU(2)_L$ and $U(1)_Y$ gauge symmetry,
respectively. That is, $T$-parity reduces the number of parameters in the model. 

Before EWSB the $T$-odd linear combination of the gauge bosons acquire a mass of order $f$,
\begin{eqnarray}
W^a_H&=&\frac{1}{\sqrt{2}}(W^a_1-W^a_2), \quad M_{W_H}= gf, \\ \nonumber
B_H&=&\frac{1}{\sqrt{2}}(B_1-B_2), \quad M_{B_H}= \frac{g^\prime\!f}{\sqrt{5}},
\end{eqnarray}
while the $T$-even linear combinations,
\be
W^a_L=\frac{1}{\sqrt{2}}(W^a+W^a_2),\quad B_L=\frac{1}{\sqrt{2}}(B_1+B_2),
\ee
remain massless and are identified with the SM gauge bosons.
After EWSB the new mass eigenstates in the neutral $T$-odd sector will be
a linear combination of the $W^3_H$ and the $B_H$ gauge bosons,
yielding an $A_H$ and $Z_H$ whose masses are
\be
M^2_{Z_H}=g^2f^2\big(1-\frac{v^2}{4f^2}\big),\qquad
M^2_{A_H}=\frac{g'^2\!f^2}{5}\big(1-\frac{5}{4}\frac{v^2}{f^2}\big).
\ee
Due to the small gauge coupling $g'$ and the factor $\sqrt{5}$
that comes from the $SU(5)$ normalization of the $U(1)$ generators,
$A_H$ is the lightest $T$-odd particle (LTOP) and could be a candidate for
(bosonic) dark matter~\cite{Hubisz:2004ft}.

\subsection{$T$-parity and fermion sector}
The Littlest Higgs model without $T$-parity contains fermionic heavy partners
only in the Top sector to engineer collective symmetry braking.
This is because only Top Yukawa coupling is large enough
to participate in collective symmetry breaking.
But all the other Yukawa couplings are sufficiently small so they need not
to participate in collective symmetry breaking.
Therefore all the Yukawa couplings except for Top Yukawa are simply
ignored in a minimal setup.

Now that the action of $T$-parity to the LHM entails doubling of charged
fermion fields in comparison with the SM. In other words,
$T$-parity interchanges a fermion doublet under $SU(2)_1$ and
a fermion doublet under $SU(2)_2$ so both go in pairs.
Furthermore, $U(1)_1$ quantum number of a fermion is identical to
$U(1)_2$ quantum number of its $T$-dual fermion.
A pair of fermions, $\psi_1$ and $\psi_2$, are imbedded into incomplete
representations $\Psi_1$ and $\Psi_2$ of $SU(5)$, respectively:
\be
\Psi_1=\left(\begin{array}{c} \psi_1 \\ 0 \\ 0\end{array}\right),\qquad
\Psi_2=\left(\begin{array}{c} 0 \\ 0 \\ \psi_2\end{array}\right),
\ee
and transform under $SU(5)$ as
$\Psi_1\to V^\ast \Psi_1$ and $\Psi_2\to V\Psi_2$, respectively.
The action of $T$-parity on the doublets takes
\be
T\,:\,\, \psi_1\leftrightarrow -\psi_2 \quad \Big(
\Psi_1\leftrightarrow -\Sigma_0 \Psi_2\Big).
\ee

As for Yukawa interactions, $\Psi_1$ and $\Psi_2$ must have the same
Yukawa couplings due to $T$-parity.    
The $T$-even linear combination, $\psi_+=(\psi_1-\psi_2)/\sqrt{2}$, becomes
a left-handed doublet under the SM $SU(2)_L$
and acquire a Dirac mass through Yukawa coupling along with
a SM right-handed singlet $\psi_R$.
The SM Yukawa interaction except for top sector is achieved
in the $T$-parity invariant way that the Yukawa couplings for $\psi_1$
and $\psi_2$ are the same as follows
\be\label{eq:Yukawa}
\frac{1}{4}\lambda_1 f \epsilon_{ijk}\epsilon_{xy}
\big[(\bar \psi_1)_i\Sigma_{jx}\Sigma_{ky}
-(\bar \psi_2\Sigma_0)_i\tilde\Sigma_{jx}\tilde\Sigma_{ky}\big]
\psi_R +h.c.
\ee
where the indices $i,j,k$ are summed over 1,2,3
and $x,y$ are summed over 4,5.

The $T$-odd linear combination, $\psi_-=(\psi_1+\psi_2)/\sqrt{2}$,
gets a large Dirac mass by allying with a right-handed mirror fermion, $\tilde\psi$,
which is imbedded into a complete representation of $SO(5)$,
\be
\Psi'=\left(\begin{array}{c} \psi\\ \chi \\ \tilde \psi\end{array}\right),
\ee
transforming as $\Psi'\to U\Psi'$, with $U$ being a global $SO(5)$ rotation.
Their masses are assumed to be $\gtrsim f$ so all of them are integrated out
at electroweak scale.
$T$-parity is acted on the mirror fermion $\Psi'$ as
\be
T\,:\,\,\Psi'\to -\Psi',
\ee
so that a Yukawa term is constructed to be invariant under $T$-parity,
\be\label{eq:lagkappa}
\frac{\kappa f}{\sqrt{2}}(\bar\Psi_2\xi\Psi'
+\bar\Psi_1\Sigma_0\Omega\xi^\dagger\Omega\Psi').
\ee
Expanding the Lagrangian in power of $1/f$, one obtains a few leading terms:
\begin{eqnarray}\label{eq:heavy}
&&\kappa f \bar\psi_-\tilde\psi + i\frac{\kappa}{\sqrt{2}}\bar\psi_-H^\dagger\chi
-i\kappa \bar\psi_+\Phi^\dagger \psi  \nonumber \\
&+&\frac{\kappa}{2\sqrt{2}f}\bar\psi_+\Phi^\dagger H^T \chi
-\frac{\kappa}{4f} \bar\psi_-(H^\dagger H+2\Phi^\dagger\Phi)\tilde\psi
-\frac{\kappa}{4f}\bar\psi_-H^\dagger H^\ast\psi
+\mathcal O(\frac{1}{f^2}),
\end{eqnarray}
in which the coupling $\kappa$ is an $\mathcal O(1)$ constant
whose precise value is sensitive to the UV physics above the UV cutoff.
The first term gives a Dirac mass to $\psi_-$ and $\tilde\psi$.
The second term allows a new quadratically divergent
contribution to Higgs mass squared, as shown in the upper diagram of 
Figure~\ref{fig:toddtop}.
The fifth term also gives rise to two such diagrams which exactly cancel out
the $T$-odd fermion one-loop divergence~(See the two lower diagrams in Figure~\ref{fig:toddtop}.).
The third term generates quadratically divergent contribution to the triplet mass
but the fifth term once again cancel out, resulting in no quadratic divergence
in triplet mass as well. Thus, we do not need to worry about quadratically divergent contributions to the Higgs ($H$ as well as $\Phi$) mass squared
from new $T$-odd fermions.
\FIGURE[l]{\epsfig{file=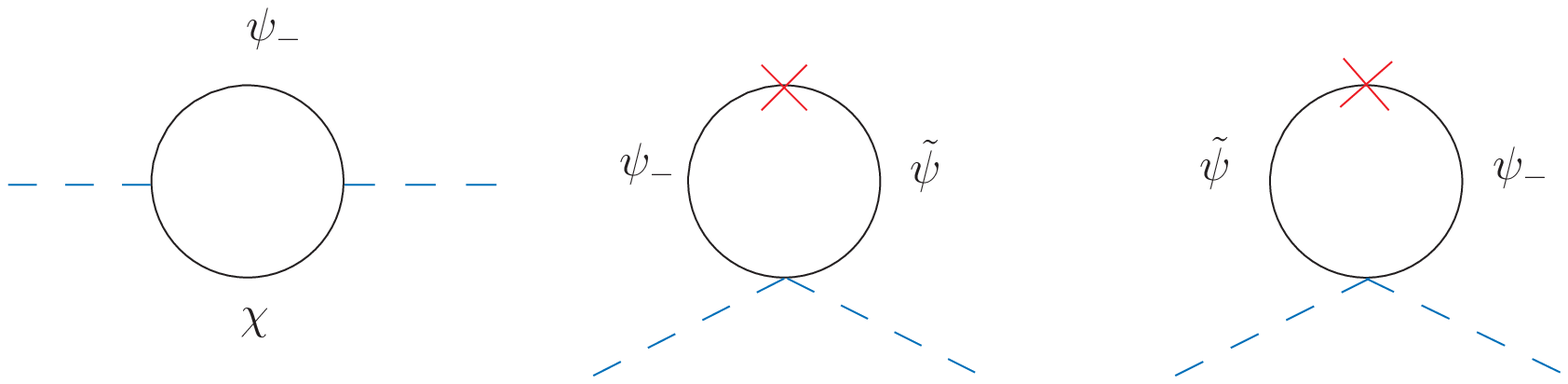,width=6in}
\caption{Contributions from the $T$-odd fermion loops to the Higgs mass parameter
cancel out.}
\label{fig:toddtop}}
Note that $\psi_+\Phi^\dagger H^T \chi$
gives no Dirac mass due to zero vev of the triplet $\Phi$.

\section{Higgs potential and $\EuScript{C}$-parity}
The Higgs fields are pNGBs in the LHM so that the Higgs potential arises from
the global symmetry breaking due to the gauge and Yukawa interactions.
The collective symmetry breaking prohibits a Higgs potential at tree level.
Instead, the Higgs potential arises at one-loop and higher orders
through the interactions with the gauge bosons and fermions. 
In addition to the collective global symmetry breaking, $T$-parity provides even
tighter constraints on the Higgs potential.
The invariance of the Higgs potential under $T$-pairty forbids $T$-odd terms.
For example, there is no tri-linear $H\Phi H$ term in the Higgs potential,
so that no mixing between $H$ and $\Phi$ take place.
In what follows, we rigorously describe the role of $T$-parity in establishing
the Higgs potential, and find an intrinsic connection between collective symmetry
breaking and $T$-parity. Thus we identify a $Z_2$ symmetry of
the LHM in the context of collective symmetry breaking.

\subsection{Gauge loop}
Let us first examine the gauge contribution to the Higgs potential.
The quadratically divergent contribution to the Coleman-Weinberg potential from gauge
boson sector is
\be\label{eq:cw1}
\frac{1}{2}a f^4 \Big\{ g^2_j\sum_b \mbox{Tr}\,[(Q^b_j \Sigma)(Q^b_j\Sigma)^\ast]
+g^{\prime2}_j\mbox{Tr}\,[(Y_j\Sigma)(Y_j\Sigma)^\ast ] \Big\}.
\ee
Here $a$ is a dimensionless coefficient which depends on the UV physics,
and the generators of the gauge groups are
\bea
Q_1^b&=&\left(\begin{array}{ccc} \sigma^b/2 & 0 & 0 \\
0 & 0 & 0 \\
0 & 0 & 0 \end{array}\right),\qquad\quad
Y_1=\mbox{diag}(3,3,-2,-2,-2)/10, \nonumber\\
Q_2^b&=&\left(\begin{array}{ccc} 0 & 0 & 0 \\
0 & 0 & 0 \\
0 & 0 & -\sigma^{b\ast}/2 \end{array}\right),\qquad
Y_2=\mbox{diag}(2,2,2,-3,-3)/10,
\eea
where $\sigma^b$ are the Pauli matrices.
Expanding $\Sigma$ in power of $1/f$,
we obtain the effective Higgs potential of the form~\cite{Han:2003wu},
\begin{eqnarray}\label{eq:gauge_coef}
V_{\rm eff}(H,\Phi)&= &\frac{a}{2}(g^2_1+g'^2_1)
\Big[f^2\mbox{Tr}\,(\Phi^\dagger\Phi)-\frac{if}{2}(H\Phi^\dagger H^T-H^\ast\Phi H^\dagger)
+\frac{1}{4}(HH^\dagger)^2+\cdots \Big] \nonumber \\
&+&\frac{a}{2}(g^2_2+g'^2_2)
\Big[f^2\mbox{Tr}\,(\Phi^\dagger\Phi)+\frac{if}{2}(H\Phi^\dagger H^T-H^\ast\Phi H^\dagger)
+\frac{1}{4}(HH^\dagger)^2+\cdots \Big].
\end{eqnarray}
Note that the signs of the tri-linear terms in the first and second lines
are opposite. This is because the $SU(2)_1\times U(1)_1$ interactions preserve
the global $SU(3)$ symmetry in the lower $3\times3$ block of $\Sigma$ while
the $SU(2)_2\times U(1)_2$ interactions do the global $SU(3)$ symmetry
in the upper $3\times3$ block of $\Sigma$.
So every interaction with an odd number of the Higgs fields
appear in pairs but with opposite signs, and thus cancels out
in the effective potential with a help of the gauge coupling conditions,
Eq.~(\ref{eq:gaugecoupling}).
As a result the effective Higgs potential consists of
interactions only with an even number of the Higgs fields:
\be\label{eq:coleman1}
V_{\rm eff}(H,\Phi)= 2a(g^2+g'^2)
\Big[f^2\mbox{Tr}\,(\Phi^\dagger\Phi)+\frac{1}{4}(HH^\dagger)^2+\cdots \Big]. 
\ee
These are a mass term for the triplet, a quartic term for the doublet
and nonrenormalizable terms which are omitted.

Invariance under the interchange of the two $SU(3)$ symmetries {\it indeed}
cancel out an odd number of the Higgs fields, and plays the same role
of $T$-parity on the Higgs sector.
Since the two $SU(3)$ symmetries are called {\it collective symmetry}
the $Z_2$ symmetry is dubbed
``$\EuScript{C}$-parity" in which $\EuScript{C}$ stands for ``collective".

\subsection{Fermion loop}
We now turn our attention to the fermion contribution to the Higgs potential,
especially in the Top sector.
Similar to $\Psi_1$ and $\Psi_2$ in Subsection 2.3,
the two fermions $Q_1$ and $Q_2$,
\be
Q_1=\left(\begin{array}{c}
q_1 \\ t_1' \\ 0
\end{array}\right),\qquad
Q_2=\left(\begin{array}{c}
0 \\ t_2' \\ q_2
\end{array}\right),
\ee
transform under $T$-parity as
\be
T\,:\,\,Q_1\leftrightarrow -\Sigma_0 Q_2,
\ee
where $q_1, q_2$ correspond to $\psi_1, \psi_2$ and $t'_1, t'_2$
are singlets under $SU(2)_{1,2}$.
$t'_i \,(i=1,2)$ are vectorlike particles along with 
additional singlet $t'_{iR}$ which transfom under $T$-parity as
\be
T\,:\,\, t'_{1R}\leftrightarrow -t'_{2R}.
\ee
The top Yukawa interaction is achieved in the $T$-parity invariant way
that not only the Yukawa couplings for $Q_1$ and $Q_2$ are the same
but also the Yukawa couplings for $t'_1$ and $t'_2$ are the same:
\be\label{eq:topYukawa}
\frac{1}{4}\lambda_1 f \epsilon_{ijk}\epsilon_{xy}
\big[(\bar Q_1)_i\Sigma_{jx}\Sigma_{ky}
-(\bar Q_2\Sigma_0)_i\tilde\Sigma_{jx}\tilde\Sigma_{ky}\big]
u_{3R} + \lambda_2f (\bar t'_1t'_{1R}+\bar t'_2t'_{2R})+h.c.
\ee
where the indices $i,j,k$ are summed over 1,2,3
and $x,y$ are summed over 4,5.
The quadratically divergent contribution to the Coleman-Weinberg potential
from the top sector is
\be
-\frac{1}{16}a' \lambda^2_1f^4\epsilon^{wx}\epsilon_{yz}
\epsilon^{ijk}\epsilon_{kmn} \big[\Sigma_{iw}\Sigma_{jx}\Sigma^{\ast my}\Sigma^{\ast nz}
+\tilde\Sigma_{iw}\tilde\Sigma_{jx}\tilde\Sigma^{\ast my}\tilde\Sigma^{\ast nz}\big],
\ee
where $a'$ is a dimensionless coefficient similar to $a$ in Eq.~(\ref{eq:gauge_coef}).
The first term arises from the $Q_1$ one-loop contribution
while the second term from the $Q_2$ one-loop contribution.
Expanding $\Sigma$ and $\tilde\Sigma$ in power of $1/f$,
we obtain the effective Higgs potential
\begin{eqnarray}\label{eq:ferm_coef}
V_{\rm eff}(H,\Phi)&= & 2a'\lambda^2_1
\Big[f^2\mbox{Tr}\,(\Phi^\dagger\Phi)-\frac{if}{2}(H\Phi^\dagger H^T-H^\ast\Phi H^\dagger)
+\frac{1}{4}(HH^\dagger)^2+\cdots \Big] \nonumber \\
&+& 2a'\lambda^2_1
\Big[f^2\mbox{Tr}\,(\Phi^\dagger\Phi)+\frac{if}{2}(H\Phi^\dagger H^T-H^\ast\Phi H^\dagger)
+\frac{1}{4}(HH^\dagger)^2+\cdots \Big].
\end{eqnarray}
The first line arises from the $\Sigma$ field contribution while the second line
from the $\tilde\Sigma$ field contribution.
Now that the expression in Eq.~(\ref{eq:ferm_coef}) is identical
with that of Eq.~(\ref{eq:gauge_coef}) except for the couplings,
we deduce that the $\Sigma$ contributions in Eq.~(\ref{eq:ferm_coef})
comes from a $SU(3)$ global symmetry in the lower $3\times3$ block of $\Sigma$
while the $\tilde\Sigma$ contributions in Eq.~(\ref{eq:ferm_coef}) from
a $SU(3)$ global symmetry in the upper $3\times3$ block of $\Sigma$.
Namely, the top Yukawa interaction, Eq.~(\ref{eq:topYukawa}), is assembled
to be invariant under $\EuScript{C}$-parity which acts on the gauge interactions.
Once again, the effective Higgs potential consists of interactions only with even
number of the Higgs fields:
\be\label{eq:coleman2}
V_{\rm eff}(H,\Phi)= 4a'\lambda^2_1
\Big[f^2\mbox{Tr}\,(\Phi^\dagger\Phi)+\frac{1}{4}(HH^\dagger)^2+\cdots \Big]. 
\ee

As to the other fermion sectors, though loop contributions from those fermions
are much smaller than those from the top sector due to their small couplings,
$\EuScript{C}$-parity still guarantees that the effective Higgs potential has
no interactions with odd number of the Higgs fields in the same way
as the top sector. Hence, $\EuScript{C}$-parity in the fermion sector
is exactly preserved.

As a last step to complete the effective Higgs potential,
we take into account the Higgs mass squared arising from logarithmically
divergent contributions. The gauge sector gives a positive Higgs mass squared
while the top sector gives a negative Higgs mass squared which dominates over
the positive gauge contribution~\footnote{Other quarks and lepton contributions
are negligible due to the Yukawa couplings.}.
As a result, the negative Higgs mass squared triggers EWSB,
\be
V_{\rm eff}=-\mu^2 H H^\dagger
\ee
with $\mu^2 >0$.
Adding the logarithmically enhanced Higgs mass squared
to the effective Higgs potential Eq.~(\ref{eq:coleman1}) and (\ref{eq:coleman2})
we finally establish the full effective Higgs potential of the form
\be
V_{\rm eff}(H,\Phi)=\lambda f^2 \mbox{Tr}\,[\Phi^\dagger\Phi]-\mu^2 H^\dagger H
+\frac{\lambda}{4}(H^\dagger H)^2+ \cdots,
\ee
where $\lambda$ is a quartic coupling and the omitted terms are 
nonrenormalizable interactions.
Due to the invariance under $\EuScript{C}$-parity the Higgs triplet
goes in pairs in the Higgs potential so its contributions to electroweak
precision constraints are in general negligible compared with those
in the original Littlest Higgs model.

\section{Origin of $\EuScript{C}$-parity}
In the previous section we have shown that $\EuScript{C}$-parity is
based on the collective global symmetry breaking.
Due to the invariance under the interchange of
the two global $SU(3)$ subgroups of the global $SU(5)$ symmetry,
particles with the same statistics are introduced so 
gauge and Yukawa interactions are established in a consistent way
that particles with the same statistics possess the {\it same} gauge 
and Yukawa couplings.
We can understand this property in analogy to supersymmetry
in which a particle and its superpartner are linked through supersymmetry,
and possess the same gauge and Yukawa couplings. But supersymmetry is broken
in nature so that masses of a particle and its superpartner are different.
$R$-parity still remains intact and indicates that their gauge and Yukawa
couplings are the same.  

Now our concerns move to a mechanism for the collective symmetry breaking.
That is, what kind of discrete symmetry corresponds to the interchange
of the upper and lower $SU(3)$ global symmetries? To answer the question
we must understand how the global $SU(5)$ is broken down to $SO(5)$
in the LHM. 
This can be achieved by assuming that the $SU(5)/SO(5)$ symmetry breaking
arises dynamically from fermion condensation through strong interactions
and the NGBs are composite particles, just as 
composite Little Higgs theories~\cite{Katz:2003sn,Lee:2005kd}.

We choose $SO(N)$ gauge symmetry for the technicolor-like strong interactions.
The global $SU(5)$ acts on the five fermions, $\varphi_2, \varphi'_2$ and $\varphi_0$,
which are dubbed ``Ultra-fermions". 
The quantum numbers of the Ultra-fermions under the gauge symmetries
are listed below.
\begin{center}
\begin{tabular}{c||c|cc|cc}
& $SO(N)$ &$SU(2)_1$ & $U(1)_1$ & $SU(2)_2$ & $U(1)_2$ \\ \hline
$\varphi_2$ & $N$ & 2 & 1/4 & 1 & 1/4 \\ 
$\varphi'_2$ & $N$ & 1 & -1/4 & 2 & -1/4\\ 
$\varphi_0$ & $N$ & 1 & 0 & 1 & 0
\end{tabular}
\end{center}
The two $U(1)$'s quantum numbers
are fixed to set $(Y_1,Y_2)=(0,0)$ for the bilinear $\varphi_2\varphi'_2$.
The bilinears of $\varphi_2, \varphi'_2$ and $\varphi_0$ correspond to
the order parameter $\Sigma$ of $SU(5)$ to $SO(5)$ breaking so the
fluctuations are the fourteen NGBs and its VEV is given as
\be
\left(\begin{array}{ccc} 
\langle \varphi_2\varphi_2 \rangle & \langle \varphi_2\varphi_0 \rangle &
\langle \varphi_2\varphi'_2 \rangle \\
\langle \varphi_2\varphi_0 \rangle & \langle \varphi_0\varphi_0 \rangle &
\langle \varphi'_2\varphi_0 \rangle \\
\langle \varphi_2\varphi'_2 \rangle & \langle \varphi'_2\varphi_0 \rangle &
\langle \varphi'_2\varphi'_2 \rangle
\end{array}\right)
=f \left(\begin{array}{ccc}
0 & 0 & 1_2 \\
0 & 1 & 0 \\
1_2 & 0 & 0
\end{array}\right),
\ee
which is proportional to $\Sigma_0$. The dimensionful parameter $f$
is assumped to be $\mathcal{O}(1)$ TeV.  
We can easily identify the fluctuations about this background
in the broken directions with the Higgs fields:
\be
\left(\begin{array}{ccc}
\Phi & H^T/\sqrt{2} & \EuScript{G} \\
H/\sqrt{2} & \EuScript{G} & H^\ast/\sqrt{2} \\
\EuScript{G} & H^\dagger/\sqrt{2} & \Phi^\dagger
\end{array}\right)=\Pi\Sigma_0,
\ee
where we explicitly include $\EuScript{G}$ which are the four
NGBs eaten by the heavy gauge fields.
We  can also identify the collective symmetries with the two
global $SU(3)$ subgroups of the $SU(5)$:
the upper $SU(3)$ acts on $\varphi_2$ and $\varphi_0$ whereas
the lower $SU(3)$ does on $\varphi'_2$ and $\varphi_0$.
Thus $\EuScript{C}$-parity is established by the invariance under
the interchange between $\varphi_2$ and $\varphi'_2$.
In other words, $\EuScript{C}$-parity is {\it naively}
realized in the NGB field matrix as
\be
\EuScript{C}\,:\,\,H\leftrightarrow H^\ast,\qquad
\Phi\leftrightarrow\Phi^\ast.
\ee 
Note that the elements in $\EuScript{G}$ are invariant under
$\EuScript{C}$-parity.
 
We now reconstruct Yukawa interactions which are invariant
under $\EuScript{C}$-parity. We get back to the the original LHM which
contains top Yukawa interactions of the form
\be\label{eq:topYu1}
\lambda_1 f \epsilon_{ijk}\epsilon_{xy}
(\bar Q_1)_i\Sigma_{jx}\Sigma_{ky}u_{3R},
\ee
where $i,j,k$ are summed over 1,2,3 and $x,y$ are summed over 4,5.
This interaction preserves the upper $SU(3)$ and
breaks the lower $SU(3)$. We need to construct the interaction
which preserves the lower $SU(3)$ and breaks the upper $SU(3)$
in a similar manner as (\ref{eq:topYu1}).
Because $\Sigma_{jx}(j=1,2,3$ and $x=4,5)$ transforms into
$\Sigma_{lv}(l=1,2$ and $v=3,4,5)$ under $\EuScript{C}$-parity 
$Q_1$ and $u_{3R}$ also transform under $\EuScript{C}$-parity,
respectively, as
\be
\EuScript{C}\,: Q_1^\dagger \leftrightarrow  Q_2,
\qquad u_{3R}\leftrightarrow u_{3R}^\dagger.
\ee
This is because the Higgs fields transform its complex conjugate
under $\EuScript{C}$-parity. 
Then the Yukawa interaction (\ref{eq:topYu1}) transforms under $\EuScript{C}$-parity as
\be\label{eq:topcdual}
\lambda_1\epsilon_{lm}\epsilon_{uvw}\bar{u}_{3R}
\Sigma_{lv}\Sigma_{mw} {(Q_2)}_u,
\ee
where $l,m$ are summed over 1,2 and $u,v,w$ are summed over 3,4,5.
Now taking the Hermitian conjugate of (\ref{eq:topcdual}) we get
\be
\lambda_1\epsilon_{lm}\epsilon_{uvw}
{(\bar Q_2)}_u\Sigma_{vl}^\ast\Sigma_{wm}^\ast u_{3R}.
\ee
Since $\Sigma$ is symmetric we can rewrite the above equation as
\be
-\lambda_1 \epsilon_{ijk}\epsilon_{xy} {(\bar{Q}_2\Sigma_0)}_i
\tilde\Sigma_{jx}\tilde\Sigma_{ky}u_{3R},
\ee
which is nothing but the $T$-dual of (\ref{eq:topYu1}).
The minus sign naturally emerges when we not only switch the running indicies 
but also replace the elements of $\Sigma^\ast$ by the elements of 
$\tilde\Sigma$ which is $T$-dual of $\Sigma$.
 
Now we are able to justify the assignment of $T$-parity on the Higgs sector, (\ref{eq:tahiggs2}).
Let us analyze the Yukawa interaction,~(\ref{eq:topYu1}),
which brings out the mass term for the SM fields.
The fermion $q_1^\dagger$ which is a doublet under $SU(2)_1$ couples to
$\Sigma_{3x}=i\sqrt{2}H/f-\Phi H^\dagger/f^2+\cdots (x=4,5)$.
The $\EuScript{C}$-dual is that the fermion $q_2$ couples to $\Sigma
_{i3}=i\sqrt{2}H^\dagger/f-H\Phi^\dagger/f^2+\cdots (i=1,2)$. Taking the hermitian
conjugate of the $\EuScript{C}$-dual is that the fermion $q_2^\dagger$ couples to
$-i\sqrt{2}H/f-\Phi H^\dagger/f^2+\cdots$. Thus the Higgs doublet couples to
$\EuScript{C}$-even fermion, $(\bar q_1-\bar q_2)$, to the leading order. The next leading
term is $-(\bar q_1+\bar q_2)\Phi H^\dagger u_R$ which shows that the triplet Higgs couples
to the $T$-odd fermion and the Higgs doublet so we can easily read off the 
transformation properties of the Higgs doublet and triplet. These transformations
are nothing but (\ref{eq:tahiggs2}) so $T$-parity is derived from $\EuScript{C}$-parity.

We have so far shown that the $T$-parity is consistent with $\EuScript{C}$-parity.
But there is a difference between $\EuScript{C}$-parity and $T$-parity.
The operators which give rise to $T$-odd fermion mass is not definitely invariant under
$\EuScript{C}$-parity. This is because the mirror fermion $\Psi'$ does not properly
transform under $\EuScript{C}$-parity. Therefore we define a new mirror
fermion $\Psi^{''}$ which transforms under $\EuScript{C}$-parity as follows
\be
\EuScript{C}\,:\,\,
\Psi''=\left(\begin{array}{c} \psi \\ \chi \\ \tilde \psi\end{array}\right)
\leftrightarrow -\left(\begin{array}{c} \tilde \psi \\ \chi \\\psi\end{array}\right)
=\tilde\Psi''.
\ee
Now we can construct the Yukawa interaction in the following $\EuScript{C}$-parity
invariant way:
\be
\frac{\kappa f}{2}(\bar \Psi_2 \xi \Psi''+ \bar \Psi_1\Sigma_0\Omega\xi^\dagger\Omega \Psi''
+\bar \Psi_2 \xi \tilde \Psi''+ \bar \Psi_1\Sigma_0\Omega\xi^\dagger\Omega \tilde\Psi'').
\ee

\section{Gauge anomalies}
A UV complete theory of the LHT should be free of gauge anomalies.
Here we consider the anomaly cancellation, in particular,
for the two $[SU(2)\times U(1)]$ gauge symmetries.
There are three kinds of fermions involved in the gauge anomalies:
({\it i}) quarks and leptons, ({\it ii}) Ultra fermions,
({\it iii}) the remaining and unknown fermions.
We do not specify the fermions in the third category in this paer.
But we will make a few remarks on the fermions in the third category
to fulfill the anomaly cancellation.

Due to $\EuScript{C}$-parity the charges of quarks and leptons are assgined
in such a way that the $U(1)_1(U(1)_2)$ quantum number of $\psi_1$ is identical
to the $U(1)_2(U(1)_1)$ quantum number of $\psi_2$, the $T$-dual of $\psi_1$.
There is a straightforward assignment of the two $U(1)$ quantum numbers
in such a way that each quark (lepton) has identical charges under the two $U(1)$'s.
Therefore a quark (lepton) has the same $U(1)$ charges with the $\EuScript{C}$-dual
partner as follows.
\begin{center}
\begin{tabular}{c|cc|cc||c|cc|cc}
 & $SU(2)_1$ & $U(1)_1$ & $SU(2)_2$  &$ U(1)_2$ & 
 & $SU(2)_1$ & $U(1)_1$ & $SU(2)_2$  &$ U(1)_2$\\ \hline
$q_1$ & $\mathbf{2}$ & 1/12 & $\mathbf{1}$ & 1/12 & 
$q_2$ & $\mathbf{1}$& 1/12 &  $\mathbf{2}$ &  1/12\\ 
$t'_1$ & $\mathbf{1}$ & 1/3 & $\mathbf{1}$ & 1/3 & 
$t'_2$ & $\mathbf{1}$ & 1/3 & $\mathbf{1}$ & 1/3\\ 
$t'_{1R}$ & $\mathbf{1}$ & 1/3 & $\mathbf{1}$ &  1/3 & 
$t'_{2R}$ & $\mathbf{1}$ & 1/3 & $\mathbf{1}$ & 1/3\\ 
$u_{3R}$ & $\mathbf{1}$ & 1/3 & $\mathbf{1}$ & 1/3 & 
$d_R$ & $\mathbf{1}$ & -1/6 & $\mathbf{1}$ & -1/6\\ 
$l_1$ & $\mathbf{2}$ & -1/4 & $\mathbf{1}$ & -1/4 & 
$l_2$ & $\mathbf{1}$ & -1/4 & $\mathbf{2}$ & -1/4\\ 
$e_R$ & $\mathbf{1}$ & -1/2 & $\mathbf{1}$ & -1/2 & & \\
\end{tabular}
\end{center}
Note that this charge assignment is different from
those in \cite{Han:2003wu,Hubisz:2004ft} because the fermions in the third category
will give a freedom for the model to be free of all gauge anomalies.
Now we can easily figure out the anomaly conditions
because we need to evaluate only the anomaly condition for $SU(2)^2_1U(1)_1$
rather than the four anomaly conditions for $SU(2)^2_{1,2}U(1)_{1,2}$.
In addition, it is enough to evaluate the anomaly condition for $U(1)_1^3$
rather than the four anomaly conditions for $U(1)^3_{1,2}$, $U(1)^2_1U(1)_2$
and $U(1)^2_2U(1)_1$.  
 
As for quarks and leptons, the anomaly conditions for $SU(2)^2U(1)$
vanishes but the anomaly conditions for $U(1)^3$ still do not.
On the other hand, the anomaly conditions for $SU(2)^2U(1)$ and $U(1)^3$
in the Ultra fermions do not vanish.
To make the model free of the guage anomalies, addition of fermions
belonging to the third category is needed. 
The standard way of obtainig no $SU(2)^2 U(1)$ and $U(1)^3$ gauge anomalies
is to add vector-like fermions to the model. 
But we will leave exploration of the full fermionic
contents in the UV complete theory for the future work.

As an aside, we would like to comment on the $T$-parity violating processes
in the context of gauge anomalies. C. Hill and R. Hill assert that the $T$-parity
violation naturally arises from the Wess-Zumino-Witten term, and
the heavy photon decays to $ZZ$ or $W^+W^-$. We can easily understand
$T$-parity violating process in the context of composite Higgs and $\EuScript{C}$-parity.
The two global $SU(3)$ subgroups of the global $SU(5)$ are explicitly broken by
gauge and Yukawa interactions. But the $T$-parity is classically conserved.
Now that quantum effects like Fig.~\ref{fig:ganomaly} breaks $T$-parity.
It is analogous to the $U(1)$ axial anomaly which explains how the decay
$\pi^0\to\gamma\gamma$ takes place. Thus the heavy photon can not be a candidate for DM.

In order to make a good DM candidate, one may take a single $U(1)$ gauge symmetry
rather than two $U(1)$ gauge symmetries in the beginning so the model
contains no heavy photon and the LTOP is a fermion rather than a boson. But in this case
the model contains an singlet NGB $\eta$ which is not eaten by the heavy photon
in case of the two $U(1)$ gauge symmetries. $\eta$ is a pNGB so that it is expected
to be light. Thus $\eta$ decays to two photons and can not be a (bosonic) DM.
Now we understand why the DM candidate must come from the strong interaction sector
as in the composite Higgs model~\cite{Katz:2003sn}.
In this regard, it is much easier to construct a model with a single $U(1)$ and
$\EuScript{C}$-parity. 

\FIGURE[l]{\epsfig{file=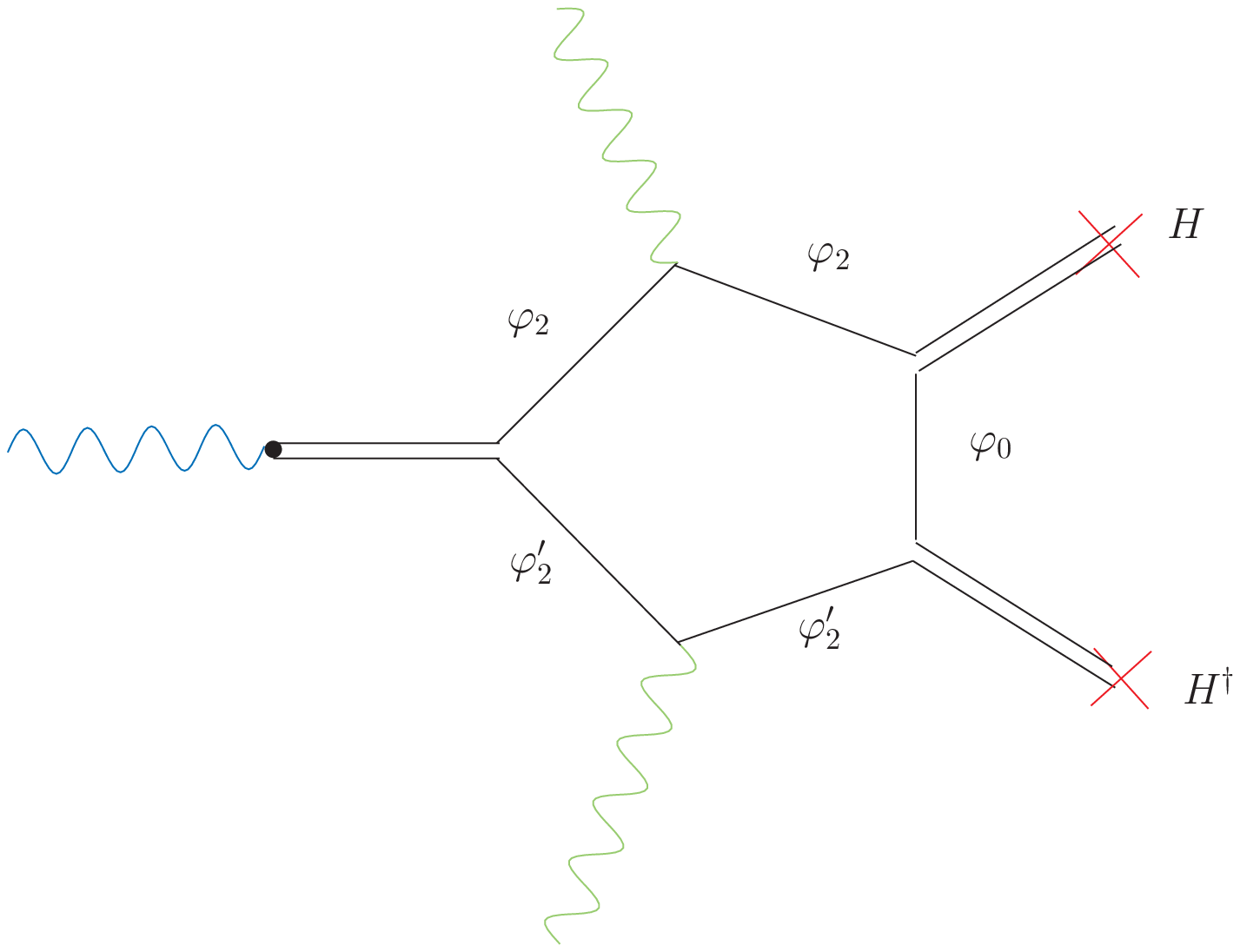,width=4in}
\caption{Gauge anomaly diagram for $A_H\to W^+W^-$ or $ZZ$.}
\label{fig:ganomaly}}
\section{Summary}
In the Littlest Higgs model the SM Higgs is a part of pNGBs arising from
a broken global symmetry and the lightness of the SM Higgs is guaranteed
by collective symmetry breaking.
To be compatible with EWPTs a $Z_2$ symmetry is introduced to the Littlest Higgs model. 
$\EuScript{C}$-parity is the $Z_2$ symmetry which makes the theories invariant 
under the interchange between the two global subgroups of the global symmetry.
Though each global subgroup is broken by both gauge and Yukawa interactions
the whole lagrangian is still invariant under $\EuScript{C}$-parity.   
$T$-parity is naturally derived by $\EuScript{C}$-parity.
$\EuScript{C}$-parity can be explained by the composite Higgs and thus provide
a guidance to a UV complete theory of the model. In addition,
$T$-parity ($\EuScript{C}$-parity) is broken at quantum level so $T$-parity
violating process is naturally understood. To make a robust DM candidate in
Little Higgs theories, we need to specify a strong interaction sector
which not only explains composite Higgs but also contains a DM parity within itself.

\section{Acknowledgements}

I would like to thank Pyungwon Ko for reviewing this paper.
I also would like to acknowledge helpful conversations with 
Dong-Won Jung and Seong Chan Park. I also express thanks to Andrew Noble
for reminding me of this old project on $T$-parity.

\appendix

\end{document}